\documentclass[conference,table]{IEEEtran}
\IEEEoverridecommandlockouts

\usepackage{cite}
\usepackage{amsmath,amssymb,amsfonts}
\usepackage{algorithmic}
\usepackage{graphicx}
\usepackage{textcomp}
\usepackage[table]{xcolor}
\usepackage{url}
\usepackage[colorlinks]{hyperref}
\usepackage{dsfont}
\usepackage{multirow}
\def\BibTeX{{\rm B\kern-.05em{\sc i\kern-.025em b}\kern-.08em
    T\kern-.1667em\lower.7ex\hbox{E}\kern-.125emX}}
\begin{document}

\title{Empirical Validation of the Independent Chip Model}

\author{\IEEEauthorblockN{Juho Kim}
\IEEEauthorblockA{Carnegie Mellon University, Pittsburgh, USA \\
juhok@andrew.cmu.edu}
\thanks{Paper category: short.}
\thanks{Our dataset is available on Zenodo: \url{https://zenodo.org/records/15556774}.}
}

\maketitle

\begin{abstract}
	The independent chip model (ICM) forms a cornerstone of all modern poker tournament strategy.
	However, despite its prominence, the ICM's performance in the real world has not been sufficiently scrutinized, especially at a large scale.
	In this paper, we introduce our new dataset of poker tournaments, consisting of results of over ten thousand events.
	Then, using this dataset, we perform two experiments as part of a large-scale empirical validation of the ICM.
	First, we verify that the ICM performs more accurately than a baseline we propose.
	Second, we obtain empirical evidence of the ICM underestimating the performances of players with larger stacks while overestimating those who are short-stacked.
	Our contributions may be useful to future researchers developing new algorithms for estimating a player's value in poker tournaments.
\end{abstract}


\section{Introduction}

Poker is a popular family of card games that has attracted attention as case studies for game theory~\cite{kuhn1951, szafronetal2013, burchetal2018}, testbenches for artificial intelligence (AI) agents~\cite{bowlingetal2017, moravciketal2017, brownandsandholm2018, brownandsandholm2019}, and social phenomena~\cite{boyd1975poker, Siler2010}.
Curiously, the mainstream poker AI literature has generally ignored the fact that, in poker, player stacks change dynamically depending on each hand's outcome. Instead, it has treated poker as a one-shot game with a static context, with identical initial states.
Such a restriction was largely for reasons related to convenience and ease of development, as treating it as a repeated game with a limited budget presents new challenges and increases the number of initial states to be considered.

This additional consideration is especially relevant in poker tournaments where players treat chips as their lifelines and try to conserve them across hands, as losing all of one's chips entails elimination.
In tournament settings, players are ranked in a reverse chronological order of their elimination, with their position determining the prize money they are eligible to win.
Typically, the prize pool is geometrically distributed such that the award decreases rapidly as one goes down the pay ladder.
As player eliminations lead to pay jumps of all other players involved, players are encouraged to play in a risk-averse manner, keeping pots small, avoiding speculative plays, refraining from aggression, being passive, and letting others clash while staying out of trouble.
This calls for a completely different strategy than in cash games, where players are (usually) allowed to buy in after losing all of their chips or top off their stacks, and this is reflected in the general poker literature, where different guides and books often frame themselves as targeting either cash games or tournaments.

Perhaps like in any other game, players' strategic considerations in poker tournaments depend on their standings relative to other players.
In poker tournaments, this is a function of one's chip count, among other things.
The \textbf{independent chip model (ICM)} is a cornerstone concept of any modern and mathematically-informed poker tournament strategy that aims to provide an estimate of their value.
The ICM takes chip count of each player and tournament payouts as inputs, and outputs nonlinear estimates of the prize money won by each player.~\cite{acevedo2019}
The algorithm can also be trivially modified to output the probabilities of each player finishing at every possible position in the tournament.

The ICM has also seen use in initializing different game-theoretic algorithms applied to poker, which found the ICM's expectations to be reasonably accurate in two-to-three player scenarios.~\cite{miltersenetal2007, ganzfriedandsandholm2008, ganzfriedandsandholm2009}
Surprisingly, despite its ubiquitous use, to our knowledge, the only peer-reviewed empirical validation of the ICM on real-world data to date was the recent small-scale analysis performed by Scott et al.~\cite{scottiiietal2024}, using 25 tournament events and 294 players.
They found that the ``ICM does a fairly good job'' but observed several discrepancies when players are grouped by stack sizes.
Gilbert~\cite{gilbert2009} mentions that Henke~\cite{henke2007}, too, found ``reasonable agreement between theory and data'' while making observations that partly disagree with Scott et al.'s. Their findings are discussed in more detail in Section~\ref{sec:background}.

In this paper, we make two contributions.
First, we release a dataset of poker tournaments scraped from \href{https://www.wsop.com/}{WSOP.com} (before their recent website redesign), yielding results of 9,958 events, payouts of 5,018 events, and end-of-day chip count summaries of 1,809 events, and \href{https://www.pokernews.com/}{PokerNews}, providing end-of-day chip count summaries of 3,610 events and payouts of 4,024 events.
Our new dataset may allow future researchers to unlock extra insights related to poker tournaments.
Second, from chip counts filtered from these two sources (2,500 relevant end-of-day chip count snapshots and results, totaling 33,478 players), we perform a large-scale empirical validation of the ICM.
In our first experiment, we compare the ICM's performance with a baseline we propose, finding that the ICM is indeed more accurate.
Next, in our second experiment, we attempt to settle the inconsistencies between the observations of Scott et al.~\cite{scottiiietal2024} and Henke~\cite{henke2007}.
We find that the ICM tends to give overly pessimistic expectations for players with larger stacks while being overly optimistic in its predictions of short stacks' performances, in line with Henke's observations.

\section{Background}
\label{sec:background}

The ICM is a canonical concept in poker that aims to approximate a player's value at any stage of poker tournaments.
While a `true' method that calculates such a value would be a function of not only player chip counts and the prize money distribution but also player positions, players' skills, blind levels, and many more, the ICM only considers chip counts and prizes as inputs.~\cite{acevedo2019}
By abstracting away the player elements of the game, the algorithm is potentially made less accurate, but its predictions are generalized so that they can apply to many different situations.

The ICM assumes that the probability of Player $i$ winning first place is linearly proportional to the percentage of chips that belong to Player $i$ (denoted as $\mathbf{x}_i$), as shown in~\eqref{eqn:first}.
\begin{equation}
	\mathds{P}(X_i = 1) = \mathbf{x}_i
	\label{eqn:first}
\end{equation}
It then assumes that the conditional probability of Player $i$ finishing second given that Player $j$ finished first is as in~\eqref{eqn:second}.
\begin{equation}
	\mathds{P}(X_i = 2 | X_j = 1) = \frac{\mathbf{x}_i}{1 - \mathbf{x}_j}
	\label{eqn:second}
\end{equation}
The conditional probability of Player $i$ finishing in $j$'th place given that Players $k_1, k_2, \hdots, k_{j - 1}$ placed first, second, and so on until $(j - 1)$'th, respectively, is expressed in~\eqref{eqn:general}.
\begin{align}
	& \mathds{P}(X_i = j | X_{k_1} = 1, X_{k_2} = 2, \hdots, X_{k_{j - 1}} = j - 1) \nonumber \\
	& \quad = \frac{\mathbf{x}_i}{1 - (\mathbf{x}_{k_1} + \mathbf{x}_{k_2} + \hdots + \mathbf{x}_{k_{j - 1}})}
	\label{eqn:general}
\end{align}
Gilbert~\cite{gilbert2009} explored risk aversion under the ICM, finding that a fair bet between two players decreases the participants' expected values while increasing those of non-participants.

In practice, implementations of the ICM exhaust through all possible payout permutations, assigning probabilities to each one, before aggregating them.
As this is very inefficient (it has a factorial time complexity), past poker programmers also developed a Monte Carlo approach to the ICM.
We consider both implementations in our analysis.

One of Scott et al.'s~\cite{scottiiietal2024} experiments represents the only peer-reviewed empirical validation of the ICM on real-world data to date, involving 25 events totaling 294 players who made it to the final day of the tournaments.
They fit a linear regression model with tournament winnings as the dependent variable and the ICM outputs as the independent variable.
Observing that the coefficient of the independent variable is $1.142$ and hence not $1$, they argued that there exists a ``significant deviation from actual outcomes.''
Then, they divided the players into three tiers depending on their chip counts, one with players in the top quartile, another with players in the middle two quartiles, and the last with players in the bottom quartile.
They observed that, on average, the first-tier players overperformed the ICM expectations, the second-tier players underperformed the ICM expectations, while the third-tier players performed very closely with the ICM expectations.

According to Gilbert~\cite{gilbert2009}, although Henke~\cite{henke2007} found ``reasonable agreement between theory and data,'' his observation of deviations differed from that of Scott et al.~\cite{scottiiietal2024}
Henke saw that ``the model modestly overestimated the probabilities of small stacks winning and modestly underestimated the probabilities of large stacks winning.''
Gilbert mentions that Henke's tests were conducted on ``World Poker Tour final tables'' but nothing more.
As Henke's Master's thesis is no longer available, it is impossible to further describe his findings beyond what Gilbert summarized.

The descriptions of Scott et al.~\cite{scottiiietal2024} and Henke~\cite{henke2007} (in addition to ours) are summarized in Table~\ref{tab:descriptions}.
Both agree that the ICM underestimates the value of larger stacks.
However, Scott et al. state that the ICM overestimates the performance of medium stacks, about which Henke makes no mention.
A contradiction can be seen in their observations for short stacks: while Scott et al. found that they perform closely with expectations, Henke found that the model was overly optimistic about their performance.
We aim to reproduce the findings of Scott et al. and Henke, albeit at a much larger scale, in an attempt to settle this unresolved debate.

\begin{table}[htbp]
	\caption{The descriptions made by Scott et al.~\cite{scottiiietal2024}, Henke~\cite{henke2007}, and us with regard to the ICM's behavior on players when grouped by stack sizes. `N/A' stands for `not available'.}
	\begin{center}
		\begin{tabular}{|c||c|c|c|}
			\hline
			\multirow{2}{*}{\textbf{Group}} & \multicolumn{3}{c|}{\textbf{The ICM Expectations}} \\
			\cline{2-4}
			& \textbf{Large stacks} & \textbf{Medium stacks} & \textbf{Small stacks} \\
			\hline
			\hline
			Scott et al.~\cite{scottiiietal2024} & \cellcolor{red} Underestimate & \cellcolor{green} Overestimate & \cellcolor{yellow} Close \\
			\hline
			Henke~\cite{henke2007} & \cellcolor{red} Underestimate & \cellcolor{gray} N/A & \cellcolor{green} Overestimate \\
			\hline
			Ours & \cellcolor{red} Underestimate & \cellcolor{yellow} Close & \cellcolor{green} Overestimate \\
			\hline
			\end{tabular}
		\label{tab:descriptions}
	\end{center}
\end{table}

\section{Methodology}

\subsection{Data}

The poker tournament dataset scraped from the web contains payouts, final results, and player chip counts at the end of each day of the tournament.
The input of the ICM is the end-of-day chip count snapshots and payouts, while the desired output of the model is the prize money won by each player (which depends on their final standing).
Several events were present in both web sources, so a deduplication step was performed.
Trivial 1-player cases that occur at each tournament's final end-of-day were filtered out since they carry no informative value.
In addition, to keep the ICM algorithm tractable, a Monte Carlo variant of the algorithm was used for end-of-day chip counts with 11 or more players (with a standard error tolerance of $0.001$, limiting the number of simulations to 10,000, and ensuring at least 100 runs are performed).
Otherwise, a classical implementation was used.
As some tournament prizes were much larger than those of others, for each event, the prize values were linearly normalized to ensure that the relevant payouts summed to one.
After preprocessing, 2,500 end-of-day chip count summaries and results, containing a total of 33,478 players, were obtained.

\subsection{Experiment 1}

There is no established algorithm for estimating a player's value in a poker tournament other than the ICM.
However, for the sake of comparison with the ICM, we propose a simple and common-sense algorithm to serve as a na\"ive baseline.
Our baseline algorithm assigns the prizes depending on the relative standing of players in chip counts, assuming that the player with the most chips finishes first, the player second-in-chips finishes second, and so on.
As this algorithm fails to recognize the magnitude of the difference in chips between players and locks each player to a single payout value (and vice versa), we expect and thus hypothesize that the ICM will perform better than this baseline.
We calculate the mean squared error between the algorithm outputs and target values as the metric, while measuring statistical significance at the 95\% confidence level with a one-sided paired t-test.

\subsection{Experiment 2}

Here, we aim to answer the question of whether the accuracy of the ICM differs depending on players' stack sizes, as observed by Scott et al.~\cite{scottiiietal2024} and Henke~\cite{henke2007}.
Following Scott et al., we group the players depending on their chip counts: large stacks (the top quartile), medium stacks (the middle quartiles), and small stacks (the bottom quartile).
We filter out situations involving more than 10 players to closely match Scott et al.'s experiment, and are left with 1,504 chip counts and results, containing 9,962 players in total.
Traditionally, the ICM has only been used in final table situations, typically involving fewer than 11 players, further justifying our design choice.
For each group, we report the mean and standard error of the ICM's residuals (differences between observations and estimates).
A significant non-zero mean would indicate that the ICM systematically gives higher (if negative) or lower (if positive) estimates than what is observed for the corresponding group.
We measure statistical significance at the 95\% confidence level with a two-sided one-sample t-test.

\subsection{Assumption}

In our t-tests, the player winnings and algorithm outputs are assumed to be independent, and the same assumption is made by Scott et al.~\cite{scottiiietal2024}
We acknowledge that, as the player values in the same event sum to one, this assumption is violated in practice.
As such, the p-values and confidence intervals would be higher than what is reported.

\section{Results}

\subsection{Experiment 1}

\begin{figure}[htbp]
	\centerline{\includegraphics[width=\linewidth]{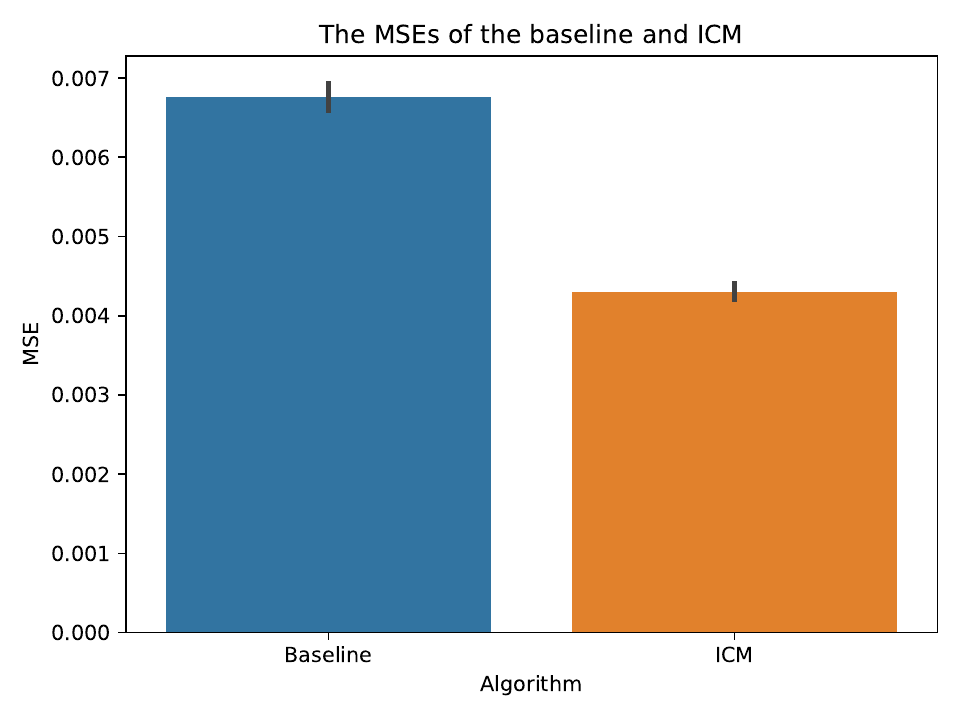}}
	\caption{The MSEs of the baseline and the ICM on the poker tournament data. Error bars indicate 95\% confidence intervals.}
	\label{fig:results}
\end{figure}

The empirical results of the two algorithms are shown in Table~\ref{tab:results} and Fig.~\ref{fig:results}.
The MSE of the baseline is $6.77 \times 10^{-3}$ (with a standard error of $9 \times 10^{-5}$), while the MSE of the ICM is $4.30 \times 10^{-3}$ (with a standard error of $5 \times 10^{-5}$).
The ICM's MSE is significantly lower than the baseline's ($p < 10^{-354}$).

\begin{table}[htbp]
	\caption{The MSEs of the baseline and the ICM on the poker tournament data. Standard errors are shown in brackets.}
	\begin{center}
		\begin{tabular}{|c||c|}
			\hline
			\textbf{Algorithm} & \textbf{MSE} \\
			\hline
			\hline
			Baseline & $6.77 \times 10^{-3}$ ($9 \times 10^{-5}$) \\
			\hline
			ICM & $4.30 \times 10^{-3}$ ($5 \times 10^{-5}$) \\
			\hline
			\end{tabular}
		\label{tab:results}
	\end{center}
\end{table}

\subsection{Experiment 2}

\begin{figure}[htbp]
	\centerline{\includegraphics[width=\linewidth]{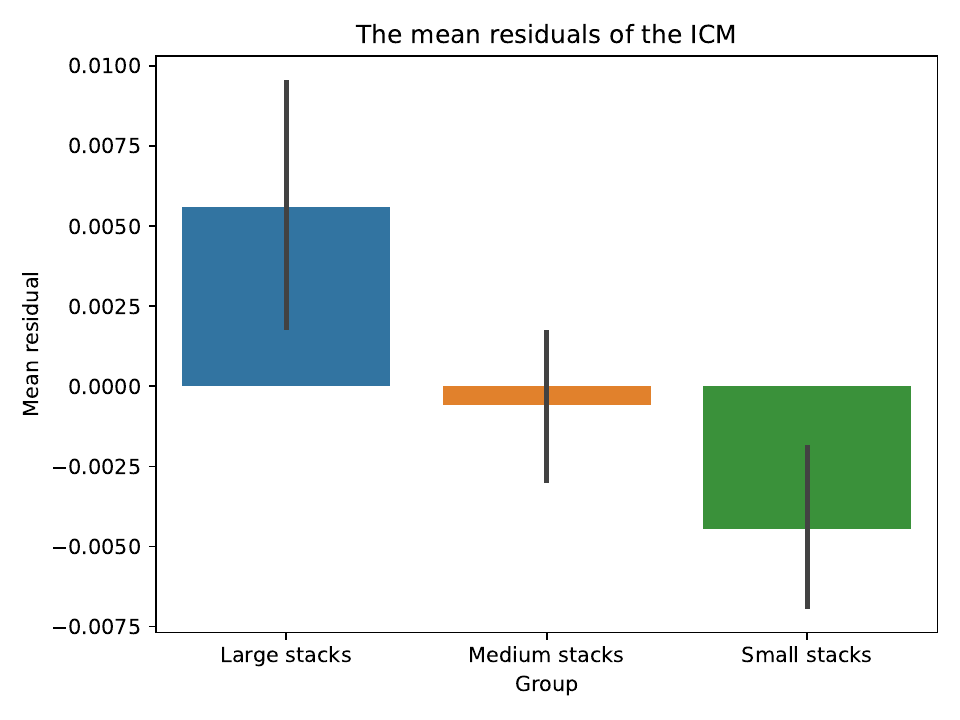}}
	\caption{The mean residuals of the ICM on the poker tournament data stratified by player stack sizes. Error bars indicate 95\% confidence intervals.}
	\label{fig:results2}
\end{figure}

The mean residual values for each group are shown in Table~\ref{tab:results2} and Fig.~\ref{fig:results2}.
Here, the mean residual for large stacks was non-zero: $5.59 \times 10^{-3}$ with a standard error of $2 \times 10^{-3}$ ($p = 0.004$).
In medium stacks, no statistically significant deviation of the mean residual from zero was observed.
For small stacks, we found that the mean residual was non-zero: $-4.44 \times 10^{-3}$ with a standard error of $1 \times 10^{-3}$ ($p < 0.001$).

\begin{table}[htbp]
	\caption{The mean residuals of the ICM on the poker tournament data stratified by player stack sizes. Standard errors are shown in brackets.}
	\begin{center}
		\begin{tabular}{|c||c|c|}
			\hline
			\textbf{Group} & \textbf{Mean residual} & \textbf{p-value} \\
			\hline
			\hline
			Large stacks & $5.59 \times 10^{-3}$ ($2 \times 10^{-3}$) & $0.004$ \\
			\hline
			Medium stacks & $-5.75 \times 10^{-4}$ ($1 \times 10^{-3}$) & $0.642$ \\
			\hline
			Small stacks & $-4.44 \times 10^{-3}$ ($1 \times 10^{-3}$) & $<0.001$ \\
			\hline
			\end{tabular}
		\label{tab:results2}
	\end{center}
\end{table}

\section{Discussion}

\subsection{Experiment 1}

In our first experiment, we found that the ICM's MSE is significantly lower than the baseline's.
In part, this difference is understandable when the design of the baseline and the ICM is considered.
One obvious room for improvement in the design of the baseline is that it commits to a single payout permutation, thus assigning each player to a single payout corresponding to their stack size.
The ICM does something similar, but instead considers every possible payout permutation, assigning a probability to each possible outcome as its weight, thus making it much more sophisticated.

\subsection{Experiment 2}

The statistically significant non-zero residual for the large stack cohort suggests that players with larger stacks tend to outperform the ICM expectations, agreeing with both Scott et al.~\cite{scottiiietal2024} and Henke~\cite{henke2007}.
Conversely, the statistically significant non-zero residual for the short stack players serves as empirical evidence that players who are short tend to do worse than what the ICM predicts.
While our observation about the short stacks agrees with Henke, it contradicts the finding of Scott et al.
The failure to establish, with statistical significance, that the residual for medium stacks is non-zero can be interpreted as the ICM giving accurate predictions of the real-world phenomena, which, again, does not align with the findings of Scott et al.
Overall, our results in the second experiment support Henke in both cases of small and large stacks, while they only align with those of Scott et al. in the case of larger stacks, as summarized in Table~\ref{tab:descriptions}.

\subsection{Future Work}

Our analysis gives clues on how a new algorithm for estimating a player's value in poker tournaments can be developed.
One can leverage multiple different algorithms like the ICM (or its variants) while selecting the most suitable one depending on the context (e.g., a player's stack size).
It may also be possible to develop an accurate algorithm by taking inputs beyond merely chip stack values, such as information about players in the tournament.
One may also develop a model that learns to rank players and assign corresponding values to them.
Regardless of what avenue is chosen, our dataset can serve as a useful resource in the training and/or evaluation of new poker tournament value estimators.

\section{Conclusion}

In this paper, we introduce our poker tournament dataset scraped from websites of two major poker tournament organizations, and carry out a large-scale empirical verification of the ICM, a ubiquitous concept in modern poker tournament theory.
We find that, first, the ICM performs more accurately than our baseline, and, second, the ICM gives overly optimistic estimates for short stack players while giving slightly pessimistic estimates for players of large stack sizes, reinforcing the findings of Henke~\cite{henke2007} in the unresolved debate between him and Scott et al.~\cite{scottiiietal2024}.
Our dataset allows new algorithms for estimating a player's value in poker tournaments to be empirically validated in a large-scale analysis, facilitating future research in the field of computer games.

\bibliographystyle{IEEEtran}
\bibliography{main}

\end{document}